\documentclass[12pt]{iopart}
\usepackage{graphicx}
\usepackage{dcolumn}
\usepackage{bm}
\usepackage{ulem}

\usepackage{color}

\begin{document}

\title{Quantum Memory with a controlled homogeneous splitting}

\author{G. H\'etet$^{1}$, D. Wilkowski$^{2,3,4}$ , T. Chaneli\`ere$^{5}$}

\address{$^1$ Institute for Experimental Physics, University of Innsbruck, A-6020 Innsbruck, Austria}
\address{$^2$ School of Physical and Mathematical Sciences, Nanyang Technological University, Singapore 637371, Singapore}
\address{$^3$ Centre for Quantum Technologies, National University of Singapore, 117543 Singapore, Singapore}
\address{$^4$ Institut Non Lin\'eaire de Nice, Universit\'e de Nice Sophia-Antipolis, CNRS, F-06560 Valbonne, France}
\address{$^5$ Laboratoire Aim\'e Cotton, CNRS UPR3321, Univ. Paris Sud, B\^atiment 505, Campus Universitaire, 91405 Orsay, France}
\begin{abstract}
We propose a quantum memory protocol where a input light field can be stored onto and released from {a single ground state} atomic ensemble by controlling dynamically the strength of an external static and homogeneous field. The technique relies on the adiabatic following of a polaritonic excitation onto a state for which {the forward} collective radiative emission is forbidden. The resemblance with the archetypal Electromagnetically-Induced-Transparency (EIT) is only formal because {no ground state coherence based} slow-light propagation is considered here. As compared to the other grand category of protocols derived from the photon-echo technique, our approach only involves a homogeneous static field. We discuss two physical situations where the effect can be observed, and show that in the limit where the excited state lifetime is longer than the storage time, the protocols are perfectly efficient and noise-free. We compare the technique to other quantum memories, and propose atomic systems where the experiment can be realized.
\end{abstract}

\maketitle

Atom-photon interfaces where photons are used as conveyers of information, and atoms are employed as stationary q-bits, are essential ingredients of quantum communication networks \cite{Cir97,DLCZ}. Besides the fundamental interest, realizing a faithful writing and read out of quantum states of light from single or collections of atoms is important for applications such as quantum repeaters, which can ensure synchronization of local operations necessary for long distance quantum communication \cite{Bri98}. Various approach are being pursued for achieving a controllable mapping between these two systems, depending on the wavelength of the light field to be stored or the material that is chosen for storage.

Efficient transfer of quantum information between atoms and photons requires controlled photon absorption with a very high probability, which can be achieved by using large atomic ensembles \cite{Pol04,Phi01}.
Reversible mapping between a weak probe and a collective atomic dark state can be achieved by adiabatically controlling in time the intensity of a control field resonant with one branch of a $\Lambda$-scheme \cite{Fle00} using the Electromagnetically-induced-transparency (EIT) phenomenon. Off-resonant Raman excitation has been also successfully considered \cite{Koz00,Dan04,NunnQM}. An alternative using a spectral hole burnt in an large inhomogeneous broadening has been proposed in the context of slow-light \cite{lauroslow}.
Another way to store quantum states of light is to employ a controlled and reversible inhomogeneous broadening (CRIB) \cite{Moi01}. Various such photon-echo type approaches were recently shown to be suitable quantum memories \cite{Sag11, Cla11, Hos11, Hed10}. Although these experiments improved rapidly over the past decades, quantum memories are still lacking long storage times, large delay-bandwidths and high efficiencies.

In this paper, we present a novel method for controllable storage of the quantum state of a light pulse in a homogeneous {single ground state} atomic ensemble via a time dependent external field. We coined this subset of quantum memories CHoS, for Controlled Homogeneous Splitting. The model is formally equivalent to EIT storage even if the slow-light propagation involves no ground state coherence or strong control fields. Here, the field state is directly mapped into optical coherences. Our approach hopefully offers a refreshing look at the concept of slow-light and extends the stopped-light experiments to a broader class of atomic systems.

We will discuss two physical situation where the CHoS can be achieved. The first memory uses the magnetically controlled dipole orientation in a V-scheme, {using a $J=0\rightarrow J=1$ transition for example}.
The second memory relies on the use of two subsets of atoms in the same medium for which one can control the detuning in an opposite fashion. This latter represents a common situation in Stark sensitive rare-earth doped materials when an homogeneous electric field is applied.
In the {weak coupling} limit, the two branches of the V-scheme indeed are independent and can thus be treated as isolated two-level systems. {In this case these two situations are formally equivalent.}
We first derive the expression of the adiabatic dark-state polariton and demonstrate the possibility to stop light in the medium when the Zeeman or Stark splitting is dynamically controlled. We then give different interpretations of the inhibited emission depending on the situation of interest. On one hand, the atomic spin polarization is rotated, and on the other hand the radiation is canceled due to a destructive interference in the forward direction. The global features of the protocol are finally derived and supported by numerical simulations. Experiments are proposed as a conclusion.

\section{Slow-light propagation}
\begin{figure}[ht]
\includegraphics[width=14cm,height=9cm]{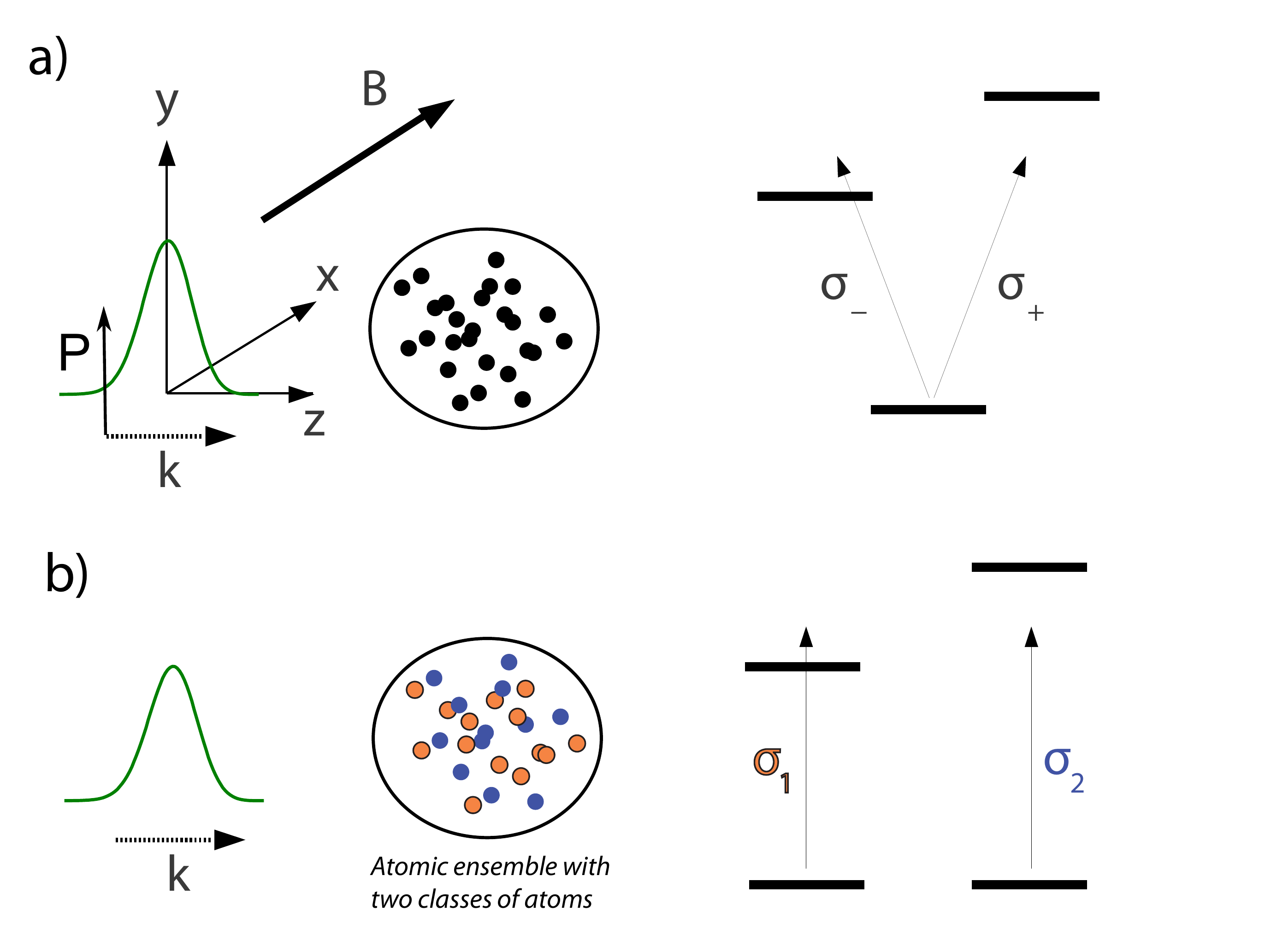}
\caption{a) The probe field is tuned between two absorbing lines (z is the propagation axis). Top: $J=0 \rightarrow J=1$ transition interacts with the circularly polarized light with respect to magnetic field axis ($x$-axis). The incoming polarization is chosen along $y$-axis so the $|F=0,m=0\rangle\langle F=1,m=0|$ {remains uncoupled throughout the probe propagation. See text for details}. b) A similar situation is obtained with two independent subsets of atoms (see section (\ref{exp}) for details). The splitting is controlled by a DC electric field (Stark effect). The common polarization orientation is here not predefined.}
\label{nivo}
\end{figure}

Our proposal follows the observation of slow-light propagation atomic ensembles and falls into a broader range of dispersive effects \cite{Boyd2002497}. Steep dispersion is automatically obtained in a transparency window surrounded by two strong absorbing features, where the absorption and dispersion interplay is imposed by the Kramers-Kroning relations. We here consider a straightforward situation where a probe field is tuned between two absorbing homogeneous lines. A dynamical control of the line splitting is then required to go from slow to stopped-light. Such a situation corresponds to the two realistic situations shown in fig. \ref{nivo}, involving Zeeman and Stark control of the transitions respectively.

Our treatment and formalism are general but we will derive explicitly both examples in order to illustrate the protocol and the corresponding physical pictures.

\subsection{Zeeman splitting case: $J=0 \rightarrow J=1$ under transverse magnetic field}

{For now}, we will assume that the atoms are excited weakly enough such the atomic population stays mostly in the ground state $|F=0\rangle$ throughout the interaction.
The quantization axis is chosen along $x$ ($B$-field). We introduce the Zeeman splitting $\Delta= g_L \mu_B B/\hbar$ ($\mu_B$ is the Bohr magneton, $g_L$ the Lande factor).

To treat the propagation of the weak field into the ensemble we follow \cite{Fle00}, and use slowly varying field envelope operators. In the circular basis, $\hat{\mathcal{E}}_y=(\hat{\mathcal{E}}_+ +\hat{\mathcal{E}}_-)/\sqrt{2}$ and $\hat{\mathcal{E}}_z=(\hat{\mathcal{E}}_+ -\hat{\mathcal{E}}_-)/\sqrt{2}i$ are the polarisation modes along $z$ and $y$ respectively. {$\hat{\mathcal{E}}_x$ is a linearly polarised field that couples to the $|F=0,m=0\rangle$ to $|F=1,m=0\rangle$}. We treat the problem in one dimension,  $z$ being the only spatial variable. {We implicitly consider atomic decay along all spatial modes in the Bloch equations. Nevertheless we only consider the coherent interaction with the sample for quantum storage purposes. We will indeed consider the coherent emission in the forward direction whose polarisation decomposes into $\hat{\mathcal{E}}_x$ and $\hat{\mathcal{E}}_y$. We thus neglect the emission in the transverse mode $\hat{\mathcal{E}}_z$ possibly excited by spontaneous process. This is justified by the incoming pulse duration assumed to be much shorter than the atomic decay time. In other words, the well-known Hanle effect consisting of the fluorescent emission polarised along $z$ is {\it de facto} neglected.}

We write $\hat{\sigma}_{\pm}^{\dagger}(z,t)$, the raising locally averaged atomic dipole operators from the ground state $|F=0,m=0\rangle$ to the excited states $|F=1,m=\pm1\rangle$, and $\hat{\sigma}_y=(\hat{\sigma}_{+}+
\hat{\sigma}_{-})/\sqrt{2}$, $\hat{\sigma}_z=(\hat{\sigma}_{+}-
\hat{\sigma}_{-})/\sqrt{2}i$.  {$\hat{\sigma}_x$ is the atomic dipole along the x-axis}.  We then write the atomic coupling strength to the two $x$ and $z$ probe modes as $g$. In the weak probe limit, we can neglect the excited-state populations and coherences {so that the Heisenberg-Langevin equations read}
$i\partial \hat{X} / \partial t=M(t)\hat{X}+\hat{\mathcal{F}}$ where
\begin{eqnarray}\label{adiab}
M=\left[\begin{array}{ccccc}kc&0&0&-g\sqrt{2N}&0\\ 0&kc&0&0&g\sqrt{2N}\\0&0&\Gamma&0&i\Delta\\-g\sqrt{2N}&0&0&\Gamma&0\\0&g\sqrt{2N}&-i\Delta&0&\Gamma\end{array}\right].
\end{eqnarray} $\hat{X}=\{\hat{\mathcal{E}}_x(k,t),\hat{\mathcal{E}}_y(k,t),\hat{\sigma}_{z},\hat{\sigma}_{x},\hat{\sigma}_y\}$
and $\hat{\mathcal{F}}$ are delta-correlated noise terms that appear due to the
fluctuation-dissipation theorem. The term $\Gamma=-i\gamma{/2}$ describes the spontaneous emission from the optical coherences. The probe detuning with respect to the central level is {set to} zero so $\Gamma$ has no real part.

{The equations of motion describing the interaction of the transverse polarization field $\hat{\mathcal{E}}_y$ with the coherences $\hat{\sigma}_y$ and $\hat{\sigma}_z$ may be derived from (\ref{adiab}) by 
Fourier transforming the optical field equation to the $z$-space}. They read

\begin{equation}\label{BMeq}
\begin{array}{rl}
\displaystyle \left(\frac{\partial}{\partial t} + c\frac{\partial}{\partial z}\right)\hat{\mathcal{E}}_y(z,t)&=-ig\sqrt{2N}\hat{\sigma}_{y}(z,t)\vspace{3mm}\\
\displaystyle \frac{\partial}{\partial t} \hat{\sigma}_{z}(z,t) &=-\frac{\gamma}{2} \hat{\sigma}_{z}(z,t)+\Delta \hat{\sigma}_{y}(z,t)+\hat{\mathcal{F}}_z\vspace{3mm}\\
\displaystyle \frac{\partial}{\partial t} \hat{\sigma}_{y}(z,t)&=-ig\sqrt{2N}\hat{\mathcal{E}}_y(z,t)-\Delta \hat{\sigma}_{z}(z,t)-\frac{\gamma}{2}  \hat{\sigma}_{y}(z,t)+\hat{\mathcal{F}}_y
\end{array}
\end{equation}
{$\hat{\mathcal{F}}_{z,y}$ are the noise terms associated with the spontaneous decays of  $\hat{\sigma}_{z,y}$.
With our choice of magnetic field and input probe polarisation directions, we note that the coherence $\sigma_x$ never gets excited so $\hat{\mathcal{E}}_x(z,t)$ does not evolve during the interaction process. It is therefore sufficient to consider the Maxwell equation for $\hat{\mathcal{E}}_y(z,t)$.}  

We now derive the formally equivalent equations of motion in the Stark splitting case.
\subsection{Stark splitting case: two subsets of atoms with opposite detunings}

We now consider two classes of atoms described by the coherence $\hat{\sigma}_{1,2}(z,t)$, which are located within a $\delta z$-thin slice  of area $\mathcal{A}$ and write
\begin{eqnarray}\label{1averaged}
\hat{\sigma}_{1,2}(z,t)&=&\frac{1}{n\mathcal{A} \delta z} \sum_{z_{k} \in \delta z} \hat{\sigma}^{k}_{1,2}(z,t),
\end{eqnarray}
The angular momentum operators of each class commute. We here assume that the transitions have opposite Stark shifts $-\Delta$ and $\Delta$ respectively. The system is governed by the same {Heisenberg-Langevin} equations that we previously derived. 

\begin{equation}\label{BMeq3}
\begin{array}{rl}
\displaystyle \left(\frac{\partial}{\partial t} + c\frac{\partial}{\partial z}\right)\hat{\mathcal{E}}_y(z,t)&=-ig\sqrt{N}\left(\hat{\sigma}_{1}(z,t)+\hat{\sigma}_{2}(z,t)\right)\vspace{3mm}\\
\displaystyle \frac{\partial}{\partial t} \hat{\sigma}_{1}(z,t) &=-\left( \frac{\gamma}{2} +i\Delta \right) \hat{\sigma}_{1}(z,t)-ig\sqrt{N} \hat{\mathcal{E}}_y(z,t)+\hat{\mathcal{F}}_{1}  \vspace{3mm} \\
\displaystyle \frac{\partial}{\partial t} \hat{\sigma}_{2}(z,t) &=-\left( \frac{\gamma}{2} -i\Delta \right) \hat{\sigma}_{2}(z,t)-ig\sqrt{N} \hat{\mathcal{E}}_y(z,t)+\hat{\mathcal{F}}_{2} \vspace{3mm}
\end{array}
\end{equation}

The analogy with the system of equations (\ref{adiab}) is complete by writing $\displaystyle \hat{\sigma}_y=(\hat{\sigma}_{2}+ \hat{\sigma}_{1})/\sqrt{2}$, $\displaystyle \hat{\sigma}_z= (\hat{\sigma}_{2}- \hat{\sigma}_{1})/\sqrt{2}i$.

\subsection{Light propagation}
For both systems, the propagation is derived from the analysis of the susceptibility that we define for the probe field mode $y$ as $\partial \tilde{\mathcal{E}}_y(z,\omega)/\partial z=\chi_y(\omega) \tilde{\mathcal{E}}_y(z,\omega)+\rm{noise}$ where $\tilde{\mathcal{E}}_y$ is the temporal Fourier transform of the probe. It reads
\begin{equation}
\chi_y(\omega)=-i\frac{\omega}{c}-\frac{\alpha}{2}\frac{\left(\gamma+i \omega\right)\gamma}{\left(\gamma+i\omega\right)^2+\Delta^2}.
\end{equation}
similar to the EIT susceptibility \cite{Fle00,PhysRevLett.66.2593} with $\Delta$ replacing the control field in the $\Lambda$-scheme.
$\alpha=4g^2N/\gamma c$ is the resonant absorption coefficient defining the optical depth of the medium $b=\alpha L$ (where $L$ is the length of the medium).
The real part of the susceptibility as a function of the probe frequency features two transmission peaks that are well separated in the limit $\Delta \gg \gamma$. The group velocity $v_g$ can now read as
\begin{equation}\label{Vg}
\frac{1}{v_g}=\frac{1}{c}+\displaystyle \frac{\alpha\gamma}{\Delta^2}.
\end{equation}
A net slow-propagation can be observed as soon as the off-resonance absorption is negligible.
The transparency condition reads as
\begin{eqnarray}
\frac{b\gamma^2}{\Delta^2}\ll1,\label{Trans_Cond}
\end{eqnarray}
The similarities with EIT will be even more flagrant by showing that stopped-light can be obtained by CHoS.

\section{CHoS dark-state polariton}

\subsection{Stopped light}\label{EIT}

Here we show that there a dark-state polariton exists \cite{Fle00} and
stopped-light can be achieved. The expression for the polariton $\hat{\Psi}$ is found by diagonalizing the matrix $M$ of Eq.~\ref{adiab}. Assuming that the probe varies smoothly in the sample so that $|k|$ is small and neglecting spontaneous decay
 yields the zero-eigenvalue (adiabatic/Dark) eigenvector
\begin{eqnarray}
\hat{\Psi}(t)=\cos \theta(t) \hat{\mathcal{E}}_y+\sin \theta(t) \hat{\sigma}_z.\label{ds}
\end{eqnarray}
The mixing angle $\theta$ is defined as
$\tan{\theta}= \frac{\alpha \gamma c}{\Delta}$ where $\theta\left(t\right)$ is now time-dependent.
Dominated by the light field, the polariton is converted into its atomic part $\hat{\sigma}_z$ when the splitting $\Delta(t)$ is reduced to zero.  After the adiabatic following, the incoming state of light is mapped into the atomic medium.
In the transparency condition, we indeed find that the group velocity is
\begin{equation}\label{vg}
v_g=c\cos^2\theta.
\end{equation}

The formal analogy with EIT-storage is nevertheless not fully satisfying.
First, the analogy is only explicit when the Autler-Townes doublet is considered in the regime of large coupling field \cite{Autler,Fle00}. In this later case, the excited state is split by the AC-Stark effect and two separate absorption lines are observed. That is the situation we achieve with a DC-Stark or Zeeman shift. Second, in the EIT case, the storage relies on the conversion of the field into the Raman coherence. In our case, no ground state coherence is involved. The equivalent of the EIT Raman coherence is here the superposition $\hat{\sigma}_z$ of the two optical coherences. The canceling of the emission therefore must rely on a different physical mechanism. In the next section we give the explanation why the splitting can dynamically control emission in the probe mode.


\subsection{Storage interpretation}

There are in fact two different reasons why the coupling to the probe mode can be canceled for the V-scheme and for the situation where two classes of atoms are considered. We explicitly separate the two cases here.

\subsubsection{Zeeman splitting or V-scheme}
At first sight, it seems surprising that the field polarization along $y$-axis is transferred to an atomic dipole along the $z$-axis, as suggested by Eq. (\ref{ds}). However, in the V-scheme considered here, the linear field polarization is decomposed into two circular polarization with opposite handedness. Each of those is then coupled to an excited state with opposite frequency detuning (see fig. \ref{nivo}a). At large Zeeman splitting, the phase shift difference due to the resonance therefore tends to $\pi$, leading to a macroscopic dipole orientation locked along $z$-axis. In this situation the medium is almost transparent and has a linear dispersion curve around the midpoint of the resonance profile. Once the magnetic field is switched off, the medium becomes opaque and the field coherence is stored in the atomic excitation. This transient effect should not be confused with the stationary regime of an on-resonance excitation (opaque) polarized along $y$-axis. There, the atomic dipole would be oriented along $y$-axis. In the protocol discussed here, the time scale is much faster than the inverse of the linewidth thus preventing a dipole reorientation. In other words, the stationary regime cannot be reached and the dipole stays along the $z$-axis. The release of the electromagnetic pulse is then performed again by simply switching on the magnetic field to its original value.


\subsection{Stark splitting case}
When two subsets of Stark detuned atoms are considered, the interpretation in terms of the previously mentioned rotated dipole emission is not appropriate anymore. Concerning the Stark shifted transition, no assumption regarding the dipole polarization was done. They are equally excited without any predefine orientation. What remains nonetheless is the destructive interference in the forward spatial mode. Tuning the splitting to zero essentially rotates the phase of the collective state so that the two classes of each atomic wavelet within one slice interferes with a relative $\pi$ phase shift. {As is clear from Eq. \ref{BMeq3}, this process transfers the coherences to $\sigma_z\propto \hat{\sigma}_{2}+e^{i\pi} \hat{\sigma}_{1}$ which is not a source field for the mode $y$. No emission can therefore take place anymore and the light field information is mapped into a spatial coherence in $\sigma_z$.}


%
%

\section{Storage performance}
\label{section_storage_performance}
Before presenting the storage simulations showing the performances of our protocol, we estimate the bandwidth and efficiency of the memory.

\subsection{Global scaling of the storage bandwidth and efficiency}\label{scaling}

This estimation has been done in the case of EIT \cite{LukinRMP} based on delay-bandwidth consideration. We simply rederive it. It has the advantage to fix the dependency between the parameters: the splitting $\Delta$, the transition width $\gamma$ and the optical depth $b$. For a given optical depth, the bandwidth should be chosen to optimize the group delay \cite{LukinRMP}. In other words, a large bandwidth requires a large splitting at the expense of a low group velocity. A trade-off should be found. On the one hand, the group delay in the slow-light limit reads as (eq. \ref{Vg}) \begin{equation}\label{T_g} \displaystyle T_g=\frac{b \gamma}{\Delta^2} \end{equation}
On the other hand, for a given optical depth, the bandwidth is not systematically given by $\Delta$ because the transparency condition (\ref{Trans_Cond}) applies.  This latter tells us that the splitting $\Delta$ should be larger than $\gamma\sqrt{b}$. When the transmission (exponential of the real part susceptibility) is plotted, $\gamma\sqrt{b}$ is the actual width of the absorption peaks. The minimum splitting is then bounded by this value $$\Delta \sim \gamma \sqrt{b}$$ imposing the condition on the bandwidth $1/\tau$ ($\tau$ is the temporal duration of the incoming pulse).
\begin{equation}\label{tau} \displaystyle \frac{1}{\tau} \sim \gamma \sqrt{b} \end{equation}

This estimation of the delay-bandwidth allows a rough estimation of the storage efficiency. Due to slow-light preparation, the pulse is spatially compressed by a factor $c/v_g$. The fraction of the incoming pulse actually contained in the medium is roughly given by $T_g/\tau$. This estimation gives the efficiency $\eta$ scaling at low optical depth.
\begin{equation} \eta \sim \displaystyle \frac{T_g}{\tau} \sim \sqrt{b} \end{equation}

Similar scaling analysis has been derived for the other storage protocols based on a steep dispersion profile. The latter may be produced by a transparency window in $\Lambda$-scheme \cite{LukinRMP}, a spectral hole burnt in an inhomogeneous profile \cite{lauroslow} or steep dispersive feature due to Raman absorption \cite{NunnQM}. Different scalings are expected depending on the nature of the dispersion profile (level structure and atom dynamics) but they all derive from a delay-bandwidth trade-off.


\subsection{Comparison with numerical simulations}


\begin{figure}[!h!]
\hspace*{0mm}\centerline{\scalebox{0.4}{\includegraphics{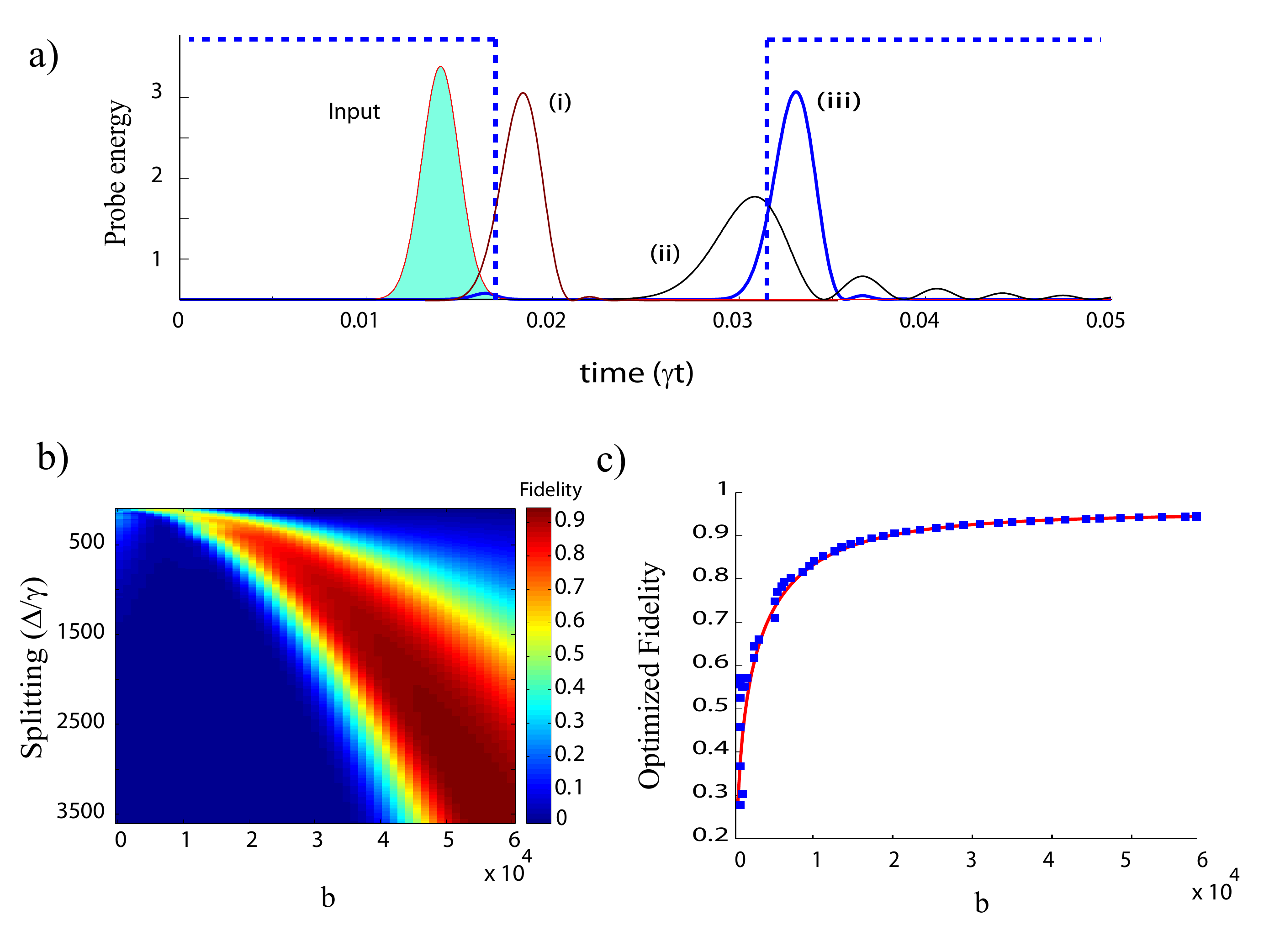}}}
\caption{a) {One dimensional Maxwell-Bloch equations simulations showing the temporal evolution of the probe intensity when the splitting is kept constant with $\Delta=3600\gamma$ and $\Delta=2000\gamma$ for traces (i) and (ii) respectively. When the splitting is dynamically controlled and initially at $\Delta=3600\gamma$ (the dashed line shows the level splitting in a.u units), the pulse is stored and retrieved efficiently (trace (iii)). b) Simulations of the evolution of the fidelity (see text for the definition) as a function of the optical depth $b$ and normalized splitting $\Delta/\gamma$. c) Optimized single mode fidelity as a function of optical depth. {It is asymptotic to 1 for large optical depth.}}}
\label{MB}
\end{figure}

To derive a more quantitative analysis {especially at large optical depth for which a good efficiency is expected}, we plotted in fig.~\ref{MB}-a) the result of numerical simulations where a weak input field is applied to an ensemble of  atoms with optical depth $b$. The pulse has a Gaussian shape with width of {$\sigma_{\tau}=0.002/ \gamma$}.

We solve the coupled Maxwell-Bloch equations including decay and probe propagation on a space-time grid for a storage time of {$0.013 / \gamma$}.
The Langevin terms associated with spontaneous emission {are not included since the atoms stay mostly in the ground state throughout the whole storage process. No noise can thus be added to the output field beyond that which ensures its commutation relations \cite{Het08, Gor07}}. Furthermore, in the weak probe limit, the atoms may be described as harmonic oscillators so that we can treat the light-atom operators as c-numbers in the simulations.

Without turning off the homogeneous splitting,
trace (i) shows a delayed output of the ensemble with a delay of $3\sigma_{\tau}$, when using {a splitting of $3600\gamma$} and an optical depth of $6.10^4$. This feature is highly demanding for an atomic sample. Nevertheless because no ground state coherence is required, we hope our proposal will stimulate the development of new materials and/or the investigation of existing media in different experimental conditions. Trace (ii) shows simulations with a {reduced splitting of $2000 \gamma$}. A larger delay is achieved but
significant distortion occurs as part of the pulse spectrum lies outside the linearly dispersive region.
In our scheme, a very high optical depth is required in order to achieve a large delay.
In order to reduce spontaneous emission, the splitting indeed has to be large enough, which lowers the group velocity, as previously discussed.

We now show that one can in fact increase the delay also by dynamical control of the transverse $B$ field intensity or Stark shift and with no distortion of the pulse shape.
Trace (iii) shows the output of the memory when the splitting is switched off at time  {$0.017/\gamma$} and on at {$0.03/\gamma$}. The output shape is here well preserved with a significant delay.
As in \cite{Fle00}, full transfer to the dark-state thus increases the delay-bandwidth that one could reach using delay only, and allows the form-stable propagation of an admixture of light and atomic coherence at a time dependent group velocity given by Eq. \ref{vg}.

We now consider the dependence of storage efficacy as function of optical depth.
To account for distortions, we calculate a single mode efficiency, which we call fidelity. It is defined as
\begin{eqnarray}
\mathcal{F}=\frac{1}{N_{ph}} \int_{t_1}^{t_2} dt \langle \hat{\mathcal{E}}^{\dagger}_{\rm out}(t-\tau)\hat{\mathcal{E}}_{\rm in}(t) \rangle,
\end{eqnarray}
where $N_{ph}=\int_{t_1}^{t_2} dt \langle \hat{\mathcal{E}}^{\dagger}_{\rm in}(t)\hat{\mathcal{E}}_{\rm in}(t) \rangle$.
The temporal mode that we use to define our detection basis is thus the input mode profile.

Fig.~\ref{MB}-b) shows the dependency of the fidelity on optical depth and splitting with a given pulse width and storage time.
The simulations show that high fidelity can be obtained only at large splitting and very large optical depths.  For low splitting values, the highest fidelity is much below 1. This is because of a strong distortion of the pulse, for which the frequency components lie outside the linearly dispersive domain. It therefore cannot be recalled efficiently in the right output temporal mode.
At higher values of the splitting, the highest efficiency increases, but higher optical depths are required. This is because the large group delay required to fit the pulse in the medium is now higher.


To find out the most favorable regime for a given input pulse, we now search for the bandwidth that maximizes the fidelity at a given optical depth. The results are shown fig.~\ref{MB}-c).
The points are the results of numerical simulations. The optimum fidelity shows a rapid increase as a function of optical depth. The red line shows analytical results using $\exp(- \gamma t_s/2)(1-\exp( \sqrt{d}/2))$ as a function, where $t_s$ is the storage time. Very good agreement is found consistent with the previous scaling predicted in sec. \ref{scaling} in the low optical depth limit.

Similar scaling analysis were derived for EIT in \cite{Gor07}. {Following the same idea, as proposed by Gorshkov \textit{et al.}, an optimized temporal shape of the control field significantly enhanced the storage efficiency. Even if it is not the goal of the present paper, it is also a possibility in our case by temporal shaping of the splitting evolution $\Delta(t)$.}


We showed here that, provided the storage time is lower than the atomic decay time, the memory can thus operate optimally at high optical depth. The decay will eventually limit the storage time but does not impact the capacity or efficiency of the memory. The latter only depends on the resonant optical depth \cite{Gor07, Nun08} which, for a homogeneous atomic ensemble, depends on the number of atoms and the scattering cross-section. Provided one can switch the magnetic or Stark shift fast, high delay-bandwidths and efficiencies can be reached. To obtain even better efficiencies if the optical depth is limited, one can mitigate coupling to another long-lived state with a detuned laser field. One could also use only a simple $\Lambda$-scheme, with an off-resonant bi-chromatic coupling field in order to create two narrow spectral lines within the probe transmission profiles. The transfer to the metastable coherence can be complete by controlling in time the frequency difference between the two components of the bi-chromatic Raman laser beam. This essentially gives similar physics then in our proposed optical coherence storage
and may complement the Raman storage protocols that were proposed with only one frequency component \cite{NunnQM}.

Let us note here that in the V-scheme storage, a longitudinal (along $z$) magnetic field would induce Faraday rotation of the probe polarization during the pulse propagation. It severely complicates the pulse propagation dynamics. However further simulations suggest that this condition becomes less drastic when storage is implemented and that the longitudinal magnetic field can in fact still be left on. {We can note also that a residual magnetic component along the polarization $y$-axis would introduce extra absorption, which may be a stringent issue at large optical depth.}



\section{Experimental proposals}\label{exp}
The storage time of CHoS memories is limited by the lifetime $\gamma/2$ of the optical coherences.  Appropriate systems should present a sufficiently long coherence time to enable switching of the magnetic or electric field required to trigger the storage.
We here focus on narrow transition in gas or in solids. Slow-light propagation have been already observed in theses systems precisely exploiting the steep dispersion produced by neighboring absorption feature \cite{Camacho, Siddons, PhysRevA.79.063844}.

\subsection{V-scheme on narrow atomic lines}

As an illustrative example on gaseous system with a natural $J=0\rightarrow J=1$ transition, we consider a cold atomic sample of strontium. The intercombination line is long-lived ($\gamma/2\simeq 2 \pi \times 7.5\textrm{kHz}$) and was already used
to show efficient and time resolved free induction decay \cite{Cha11}, a first steps towards coherent control of the collective dipole radiation with a homogeneous magnetic field. Taking some experimentally achievable quantities $\alpha=2\times10^6\,\textrm{m}^{-1}$, $\Delta=23\gamma$ corresponding to a reasonable field of $0.2\,G$ and a sample size of $L=100\,\mu$m, one gets $v_g\simeq 100\,\textrm{m/s}^{-1}$ and a pulse duration $\tau\simeq 1\,\mu$s. The transfer fidelity is expect to be around $20\%$ for $b \simeq 200$.

As previously discussed, inhomogeneous magnetic fields along $z$ should not impact the writing and reading efficiency. This should also be true for any inhomogeneous longitudinal broadening, so the method would essentially be Doppler-free. Only inhomogeneity of the B-field along the transverse mode will impact the storage efficiency resulting in a imperfect extinction of the coupling constant in the probe mode.

\subsection{Stark splitting with rare-earth dopants}

This memory proposal will here find direct use in rare-earth doped materials.
The situation where two subsets of ions exhibit opposite Stark shifts is surprisingly common in pratice. Indeed, when rare-earth ions occupy a local non-centrosymmetric substitution site embeded in matrix with overall central symmetry, the inversion operator requires the existence of opposite Stark shifts for different ions \cite{Macfarlane2007156}. Also known as pseudo-Stark spilliting, this situation has been particularly helpfull to study the surrounding of luminescent emitters in insulators \cite{Kaplyanskii200221}. Many oxyde crystals satisfy this condition \cite{Graf}.

Rare-earth crystals are also known for their extremely narrow resonances at low temperature \cite{liu2005spectroscopic}. Combined with a large absorption coefficient, slow-light propagation has been observed in a transparency window generated by spectral hole-burning \cite{PhysRevA.79.063844}.
The Stark control of the group velocity is placed in this lineage and seems to be directly accessible experimentally. The possibility offered by DC electric field has been considered in a modified and particularly successful version of the photon-echo technique \cite{Hedges2010} and has been used to inhibit and trigger the emission of a rare-earth ensemble \cite{Hyper}.

Praseodymium-doped Y$_{2}$SiO$_{5}$ is well suited for our application because it exhibits a pseudo-Stark splitting. Extremely narrow absorbing peaks can be accurately prepared. Hedges et al. \cite{Hedges2010} have indeed shown that 140 dB attenuation feature can be rapidly controlled by Stark fields with a possible effective $\gamma/2\simeq 2 \pi \times 12\textrm{kHz}$. It corresponds to an optical depth $b \simeq 32$ sufficient for a proof-of-principle demonstration. Highly doped or even stoichiometric rare-earth doped sample can be considered to obtain a larger optical depth. As compared to EIT storage in atomic vapors where four-wave mixing, absorption of the control beam and collisions were pointed out as limitations at large optical depth \cite{PhysRevA.78.023801}, our protocol combined with the specificities of rare-earth crystals may present optimized storage performances.

\section{Conclusion}

In conclusion, we propose a quantum memory protocol where either control over the three-dimensional angular distribution of the macroscopic dipole emission or the tuning of the phase difference between the atom and the field in a given mode is used. Although the physics of our storage process relies on time-encoding of the information onto dark-state and slow-light effects, it
significantly differs in essence from the EIT and Raman processes because no ground state coherence is necessary. Numerical and analytical analysis shows that the method is ideally efficient and allows large delay-bandwidth to be reached.

We here considered an optical thick ensemble, but the technique can be implemented as well with single atoms in high-finesse optical cavities. Changing the transverse magnetic field would effectively tune the strength of the Purcell effect that takes place along the cavity mode, thereby also offering a realistic prospect for single atoms quantum memories \cite{Spe11} without strong control fields. We note that a theoretical idea was also recently put forward to store light by controlling the coupling constant of rare-earth ions \cite{Gre11} and that switching the direction of magnetic field of hyperfine levels was also investigated in the context of nuclear resonant forward scattering of X-rays in \cite{Shv99}.
We anticipate that the simplicity of our idea will open a new direction for research on atom-light interfaces in a broader range of atomic systems.

\section*{Acknowledgments}
G.H would like to acknowledge useful discussions with H. Zoubi and M. Hennrich as well as financial support from a Marie Curie Intra-European Action of the European Union, the FP7 QuRep project, and by the french national grant ANR-09-BLAN-0333-03. The research leading to these results has received funding from the People Programme (Marie Curie Actions) of the European Union's Seventh Framework Programme FP7/2007-2013/ under REA grant agreement no. 287252. {We thank the National University of Singapore CQT for supporting the visit of G.H. and T.C.}

\section*{References}
\bibliographystyle{iopart-num}
\bibliography{chos,EIT,StarkPr}

\end{document}